\documentclass[12pt,preprint]{aastex}

\newcommand{\vis}{\mathcal{V}}
\newcommand{\ud}{\theta_{\rm UD}}
\newcommand{\ld}{\theta_{\rm LD}}

\shorttitle{The Interferometric Radii of Low-Mass Stars}
\shortauthors{Berger et al.}

\begin{document}

\title{First Results from the CHARA Array. IV.  
       The Interferometric Radii of Low-Mass Stars}

\author{D. H. Berger\altaffilmark{1}}
\affil{The CHARA Array, Mount Wilson Observatory, Mount Wilson, CA 91023}
\email{berger@chara-array.org}

\author{D. R. Gies, H. A. McAlister}
\affil{Center for High Angular Resolution Astronomy, Georgia State University, 
	P.O. Box 3969, Atlanta, GA 30302-3969}
\email{gies@chara.gsu.edu, hal@chara.gsu.edu}

\author{T. A. ten Brummelaar}
\affil{The CHARA Array, Mt. Wilson Observatory, Mt. Wilson, CA 91023}
\email{theo@chara-array.org}

\author{T. J. Henry}
\affil{Center for High Angular Resolution Astronomy, Georgia State University, 
	P.O. Box 3969, Atlanta, GA 30302-3969}
\email{thenry@chara.gsu.edu}

\author{J. Sturmann, L. Sturmann, N. H. Turner}
\affil{The CHARA Array, Mt. Wilson Observatory, Mt. Wilson, CA 91023}
\email{judit@chara-array.org, sturmann@chara-array.org, nils@chara-array.org}

\author{S. T. Ridgway, J. P. Aufdenberg}
\affil{National Optical Astronomy Observatory, P.O. Box 26732, Tucson, 
	AZ 85726}
\email{ridgway@noao.edu, jasona@noao.edu}

\and

\author{A. M\'erand}
\affil{LESIA, UMR8109, Observatoire de Paris-Meudon, 5 place Jules Janssen, 92195 Meudon Cedex, France}
\email{antoine.merand@obspm.fr}

\altaffiltext{1}{present address:  University of Michigan, Dept.~of Astronomy,
500 Church St., 917 Dennison Bldg., Ann Arbor, MI 48109-1042}

\begin{abstract}
We have measured the angular diameters of six M dwarfs with the CHARA Array, a
long-baseline optical interferometer located at Mount Wilson Observatory.
Spectral types range from M1.0 V to M3.0 V and linear radii from 0.38 to
0.69 $R_\sun$.  These results are consistent with the seven other M-dwarf radii
measurements from optical interferometry and with those for sixteen stars in
eclipsing binary systems.  We compare all directly measured M dwarf radii to
model predictions and find that current models underestimate the true stellar
radii by up to 15-20\%.  The differences are small among the metal-poor
stars but become significantly larger with increasing metallicity.  This
suggests that theoretical models for low mass stars may be missing some opacity
source that alters the computed stellar radii.
\end{abstract}

\setcounter{footnote}{1}


\section{Introduction}
\label{sect-intro}

Cool, low-mass stars dominate the stellar census \citep[]{Henry97,Reid04},
yet they remain elusive and enigmatic objects.  Because of their small size and
cool surface temperature, nearby members of our solar neighborhood are still
being discovered via proper motion surveys
\citep[]{Hambly04,Lepine05,Subasavage05} and parallaxes are being determined
for these new stellar neighbors \citep[]{Jao05, Costa05}.  Furthermore, the
low-mass stars we do know about are not well understood.  Their fundamental
properties are difficult to measure and do not adequately constrain atmospheric
and interior stellar models.  In addition to effective temperature and mass,
the size of field stars at the cool end of the main sequence is arguably one of
the most difficult properties to determine.
 
There are currently only two methods to measure directly the stellar radii of
cool dwarfs: light curve and radial velocity studies of double-lined eclipsing
binaries and long-baseline interferometry of single stars.  The former method
is biased toward main sequence stars larger than the Sun \citep[]{Andersen91},
and the latter toward brighter and larger giants and supergiants.  Within the
past ten years, cooler and smaller stars are being added to the database of
stars with known fundamental properties.  There are sixteen known M-dwarfs that
are members of eclipsing binaries for which we have have stellar radii:  both
components of GJ 2069A \citep[]{Delfosse99}, both components of CM Dra
\citep[]{Lacy77, Metcalfe96}, both components of CU Cnc \citep[]{Ribas03}, both
components of YY Gem and one component of V818 Tau \citep[]{Torres02}, both
components of BW3 V38 \citep[]{Maceroni04}, one component of RXJ2130.6+4710
\citep[]{Maxted04}, both components of TrES­Her0-07621 \citep[]{Creevey05}, and
both components of GU Boo \citep[]{LP05}.  From long-baseline optical
interferometry, the situation is more bleak as there heretofore were only seven
M dwarfs (GJ 15A, GJ 191, GJ 205, GJ 411, GJ 551, GJ 699, and GJ 997) for which
stellar radii have been measured \citep[]{Lane01,Segransan03}.

Here we report on measurements of the angular diameters of six M dwarfs.
We obtained these measurements from observations made with the Center for High
Angular Resolution Astronomy (CHARA) Array, a 6-element optical/near-infrared
interferometer located at Mount Wilson Observatory, California.  From these
measurements, we deduce the linear sizes and further refine the mass--radius
relation.  In addition, we calculate their surface gravities and effective
temperatures.  This paper is the fourth in a series of commissioning science
observations with the CHARA Array.  The other three papers vary in topic from
the rapid rotators Regulus \citep[]{McAlister05} and Alderamin
\citep[]{vanBelle06} to an overview of the CHARA Array \citep{Theo05}.

\section{Observations}
\label{sect:observations}

The majority of observations were completed in 2004 June, while others were
mixed into the standard queue observing throughout the end of 2003 and most of
2004.  All observations were made using the $K'$ filter ($\lambda_0 =
2.13~\mu$m, $\Delta\lambda = 0.35~\mu$m; note that the central wavelength for
the filter alone differs slightly from the system effective wavelength adopted
by \citealt{McAlister05}, but the difference has a negligible effect on our
results).  Data for GJ 15A were obtained using the most western and eastern
telescopes (W1 and E1, respectively) and for GJ 880 using the inner southern
telescope (S2) and W1.  The data for the remaining targets were observed with
the most southern telescope (S1) and E1.  The maximum baseline separations
between these telescopes are 314 meters for W1-E1, 249 meters for S2-W1, and
331 meters for S1-E1.  In these configurations, angular diameters as small as
0.5 milliarcsecond (mas) can readily be measured at the $K'$ spectral band.  

Targets were chosen from the Gliese and Jahrei\ss ~catalog \citep[]{Gliese91}
based on color ($B-V>1$) and distance (parallax $\pi>100$ mas) such that their
predicted angular size exceeded 0.4 mas.  Targets also had to be within the
detection limits of the instrument ($B<11$ for tip-tilt correction, $V<10$ for
image acquisition, and $K'<6$ for the near-IR detector).  Table~\ref{tbl-obs}
is a summary of the observations for this paper.  Column 1 is the Gliese and
Jahrei\ss ~catalog designation, column 2 is the Luyten Half-Second (LHS)
Catalog designation \citep[]{Luyten79}, column 3 gives the name of the
calibrator star, column 4 indicates the telescope pair (and baseline) used,
column 5 gives the UT observation date, and column 6 gives the number of
observations.

Measurements of all but GJ 15A were taken using the ``CHARA Classic'' beam
combiner.  It is a two-beam, pupil-plane (or Michelson) combiner utilizing path
length modulation; details of the instrumentation and configuration are given
by \citet[]{Judit03}.  GJ 15A was observed with FLUOR, a single-mode fiber beam
combiner designed by collaborators at the Paris Observatory \citep[]{FLUOR2},
because it was the available instrument at the time and the target was within
its sensitivity limits.  The calibrator for GJ 15A was HD 2952, and was chosen
from the ``Calibrator Stars for 200m Baseline
Interferometry''\footnote{http://vizier.hia.nrc.ca/viz-bin/Cat?J/A\%2bA/433/1155}
 catalog \citep[]{Merand05}, which includes corrections for limb darkening.
While FLUOR benefits from spatially cleaned beams, the ``CHARA Classic'' beam
combiner has greater sensitivity.  Hence the two instruments complement each
other in this respect.  

The same near-IR detector was at the back end of each beam combiner.  The
fringe sampling frequency was either 100 or 150 Hz, depending on the seeing
conditions and source brightness.  For the same reasons, either 1 or $2\times2$
pixels were read out.  The camera readout was adjusted to maintain five samples
per fringe.  Each merged data scan from ``CHARA Classic'' was formed from 200
scans with photometric calibration scans made before and after.  Photometric
calibration scans for FLUOR are performed during the scan on separate fiber
channels.

Atmospheric and instrumental coherence losses were estimated by interleaving
measurements of unresolved stars or stars with known angular diameters.
Calibrators were chosen using the {\it gcWeb} utility available online from the
Michelson Science
Center\footnote{http://mscweb.ipac.caltech.edu/gcWeb/gcWeb.jsp}.  We restricted
our selection to main sequence stars with estimated angular sizes less than 0.4
mas, which yields visibility amplitudes greater than 90\% when measured with
the S1-E1 baseline in the $K'$-band.  The calibrator observations were in close
proximity on the sky (within $10^\circ$) and in time (within 30 minutes) to
each source observation.  The typical duty cycle from the start of one
observation to the start of the next was approximately 10 minutes.  

\section{Data Reduction}

The reduction algorithms developed for data from the CHARA Array are described
in detail by \citet[]{Theo05}.  We employed a commonly used technique of
integrating the power of each fringe scan \citep[]{Benson95}, but we removed
the variance term resulting from atmospheric turbulence.  The resulting
quantity is the visibility amplitude ($\vis$), and not the more common
$\vis^2$.  Data from GJ 15A and its calibrator were analyzed by methods
specific to the FLUOR instrument \citep[]{FLUOR,Perrin03} and the following
description does not pertain.

The calibrated visibility amplitude of the science object ($\vis_o$) was
calculated using the relation
\begin{equation}
\vis_o=\frac{1}{\eta} \vis_o'
\label{eq:cal_vis}
\end{equation}
where $\eta$ is the interferometer's efficiency and $\vis_o'$ is the
instrumental visibility amplitude of the science object as measured on the sky.
By interleaving measurements of a calibrator of known angular size and thus a
known calibrated visibility amplitude, one can measure $\eta$ via the ratio of
the instrumental to calibrated visibility amplitude, or $\eta =
\vis'_c/\vis_c$.  Furthermore, if the calibrator is unresolved, $\vis_c$ is
unity, and $\eta$ is measured directly.  However, the high resolution at long
baselines and instrument sensitivity limitations significantly reduce the
number of available unresolved sources, especially within close proximity to
the science object.  Therefore, we were left to use slightly resolved
calibrators and to determine $\vis_c$ through other means.  

\subsection{Angular Diameters of the Calibrators} 
\label{sect-irfm}

From conservation of energy, the true limb-darkened angular diameter of a star
($\ld$) is related to ratio of the stellar flux reaching the top of the Earth's
atmosphere ($F^\earth_\lambda$) at wavelength $\lambda$ to the flux leaving
the stellar surface ($F^\star_\lambda$):
\begin{equation}
\frac{\ld^2}{4} = \frac{F^\earth_\lambda}{F^\star_\lambda}.
\label{eq:coe}
\end{equation}
$F^\star_\lambda$ is determined from stellar model atmospheres and
$F^\earth_\lambda$ from extinction-corrected photometry and absolute
spectrophotmetry.  This was an idea first proposed by \citet[]{Gray67} and
later refined by \citet[]{Blackwell77}.  A potential source of error in
$F^\star_\lambda$ is the uncertainty in determining the correct input
parameters for the model atmosphere.  For example, if the effective temperature
($T_{\rm eff}$), metallicity ([Fe/H]), and/or surface gravity ($\log g$) are
not well known, there can be difficulty in choosing the correct model.  As
pointed out by \citeauthor[]{Blackwell77}, the infrared region is less
sensitive to model input approximations than the visible --- hence the method
to derive angular diameters from infrared photometry is known as the InfraRed
Flux (IRF) method.  Because our observations were obtained in the $K'$-band,
our derived radii measurements are relatively unaffected by problems related to
limb darkening and/or absorption line contamination.

\begin{figure}
\begin{center}
\includegraphics[scale=0.3]{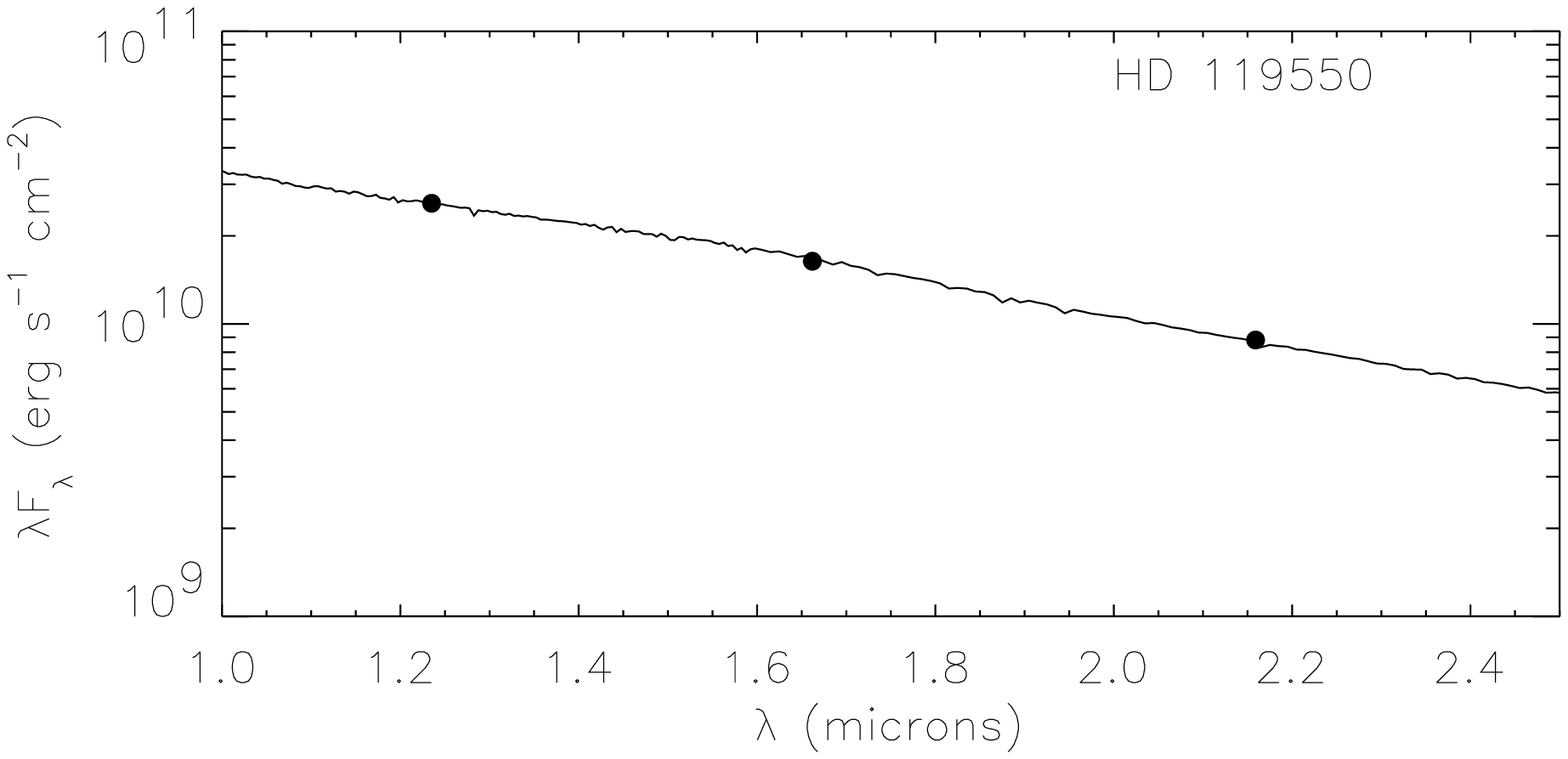}
\includegraphics[scale=0.3]{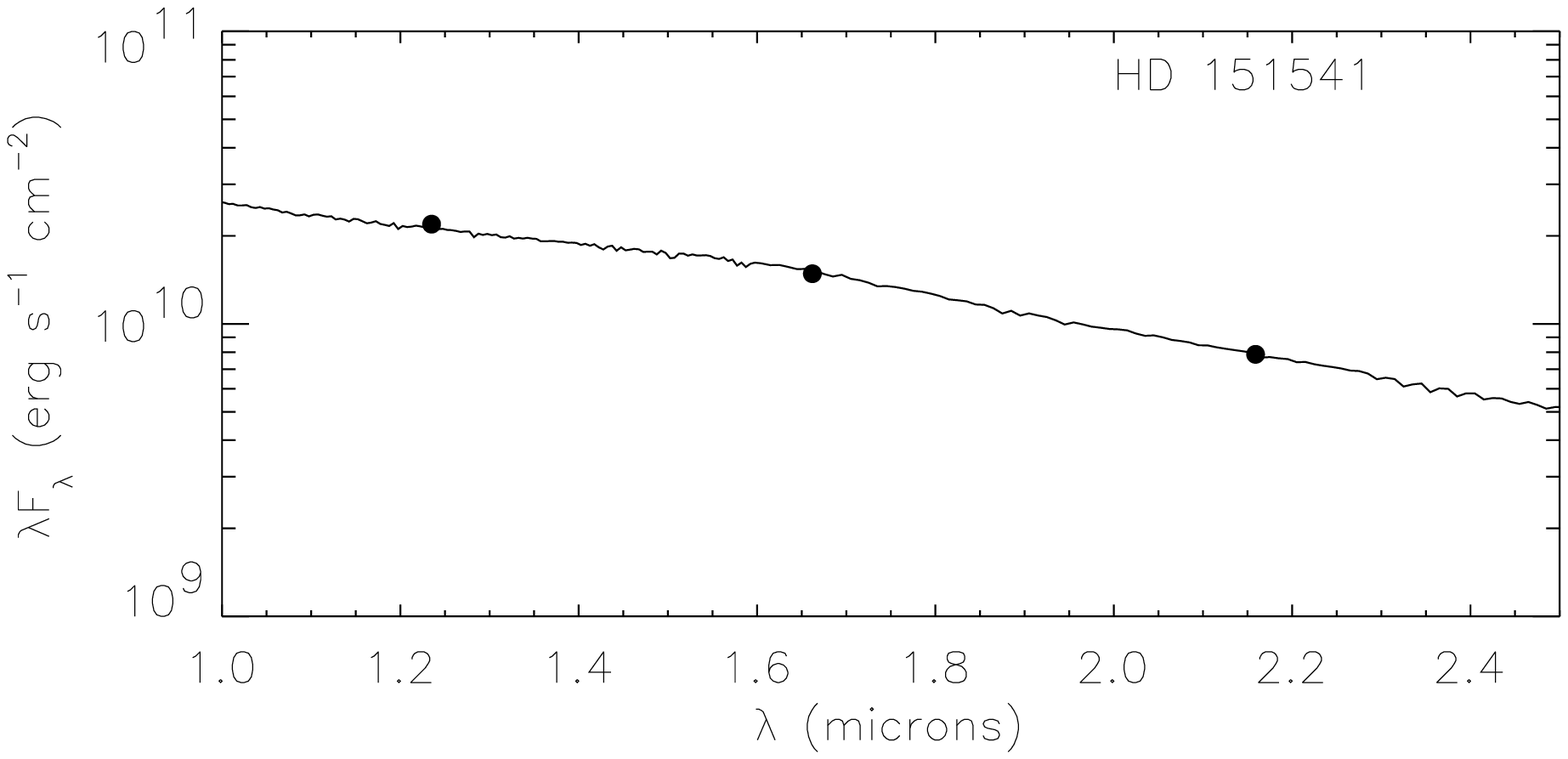}
\includegraphics[scale=0.3]{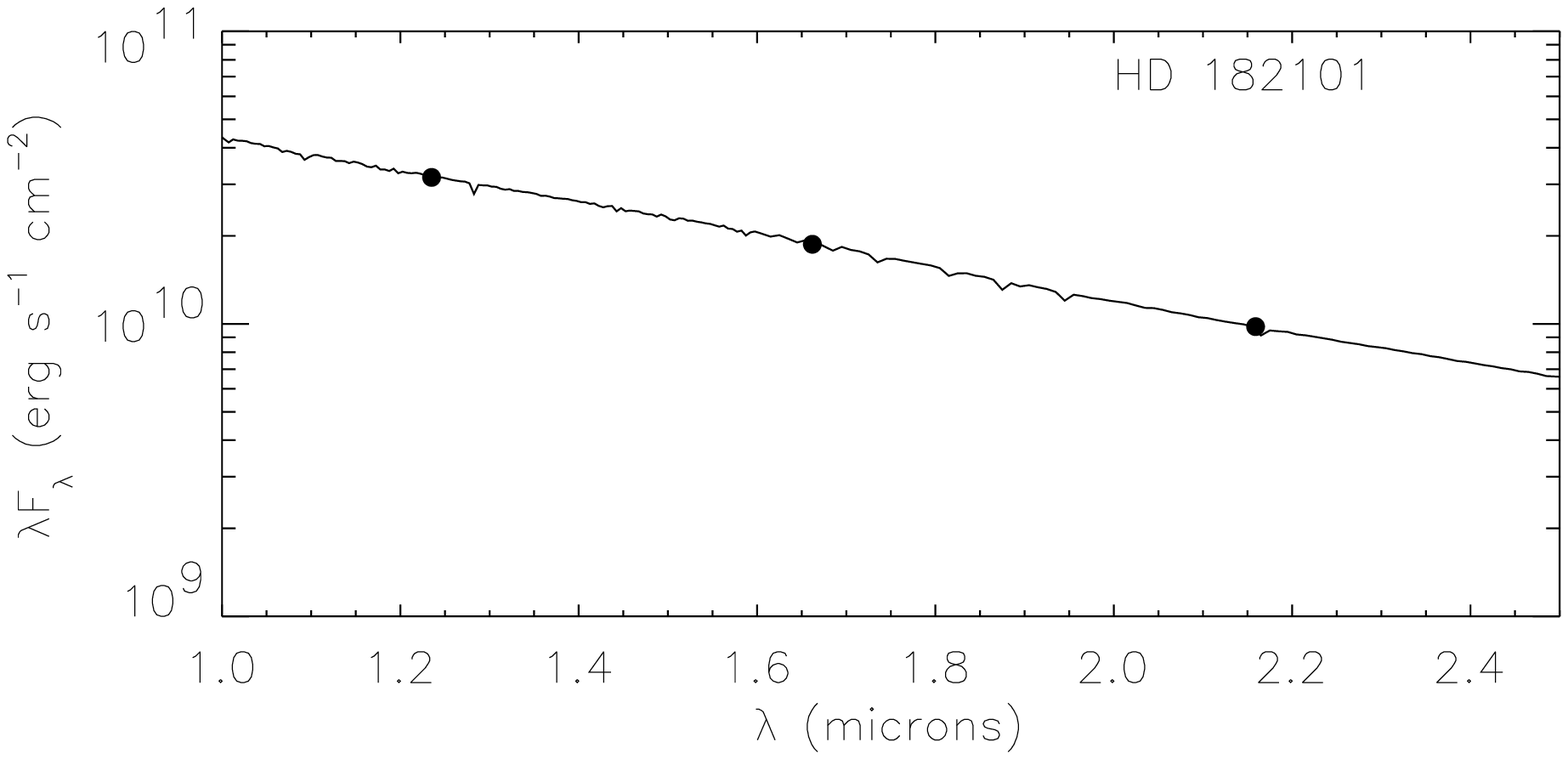}
\includegraphics[scale=0.3]{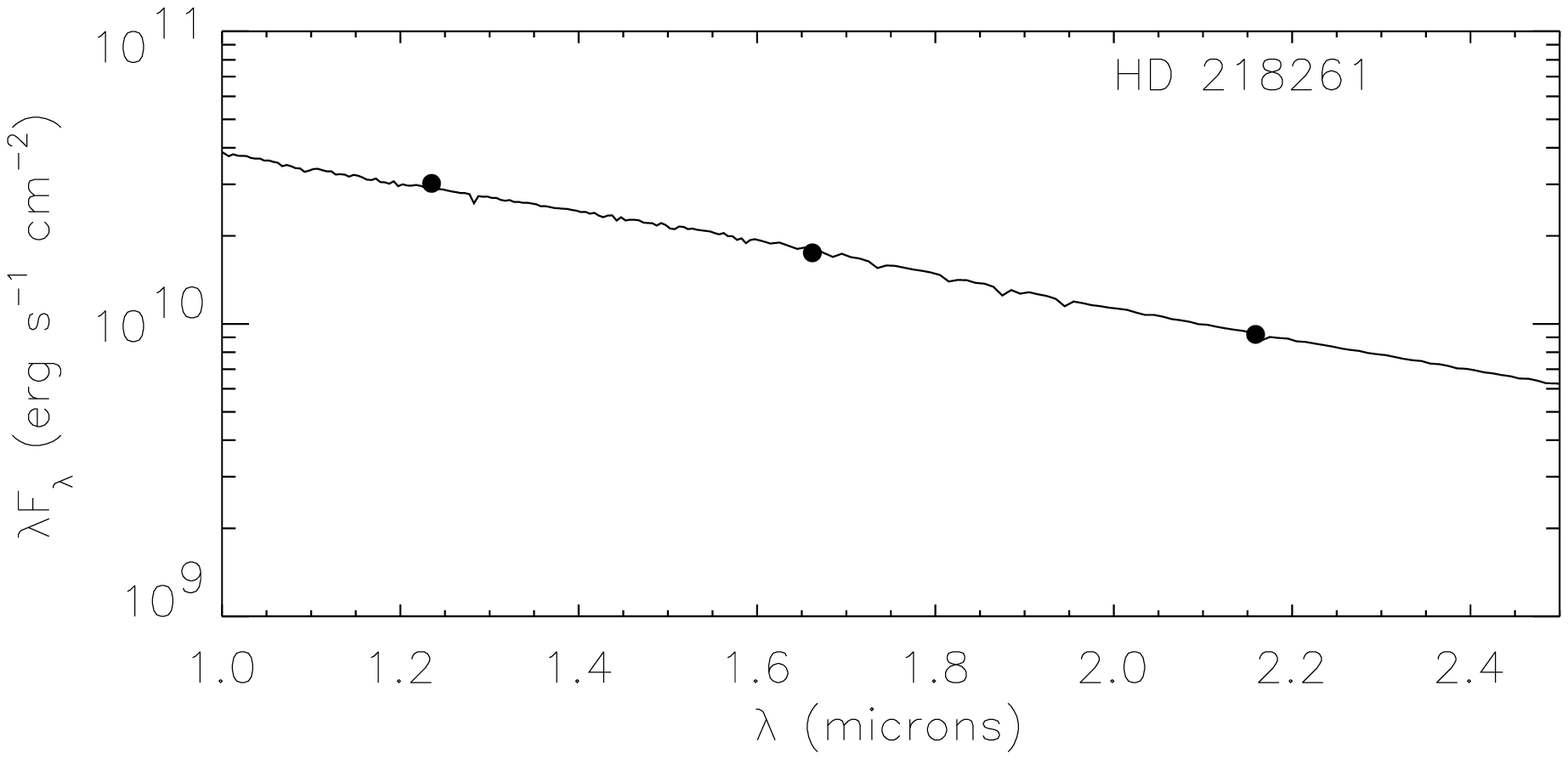}
\end{center}
\caption{Spectral energy distribution fits used to derive the
calibrator angular diameters from the IRF method.  The model atmosphere fluxes
\citep[]{Kurucz92} are shown as a solid line, and the filled circles give the
observed fluxes from the 2MASS $J,H,K$ photometry \citep[]{Cutri03, Cohen03}.
The errors in the 2MASS fluxes are approximately $\pm2\%$ (much smaller than
the symbol size).\label{fig-irf}}
\end{figure}

We selected model atmospheres from \citet[]{Kurucz92} and linearly interpolated
the fluxes between $T_{\rm eff}$ grid points.  We adopted $\log g = 4.5$, which
is appropriate for mid-temperature dwarf stars, and assumed the metallicity to
be solar.  However, changing $\log g$ and/or the metallicity did not
significantly effect the resulting $\ld$.  The Two Micron All Sky Survey
(2MASS) point source catalog \citep[]{Cutri03} provided the near-IR broad-band
photometry data, which were then transformed to fluxes \citep[]{Cohen03}.  A
weighted fit of the 2MASS photometry to the model flux yielded angular
diameters less than 0.4 mas and typical errors of 4-5\% (Figure~\ref{fig-irf}).
For calibrators with such small angular size, the propagated fractional error
in the target angular diameter is much lower than this percentage
\citep[]{vanBelle05b}.  We list the names and adopted parameters of the
calibrator stars in Table~\ref{tbl-cal}, and our derived $\ld$ values are given
there in column 5.

The IRF method yields $\ld$.  However, it is common practice to fit a
visibility function for a uniform disk angular diameter ($\ud$), which is given
by
\begin{equation}
\vis = \left|\frac{2 J_1\left(\pi B \ud/\lambda_0\right)}{\pi B
\ud/\lambda_0}\right|
\label{eq:vis_ud}
\end{equation}
and depends on the projected baseline ($B$) and the effective wavelength
($\lambda_0$).  In order to compute the efficiency ($\eta$) from the measured
visibility of the calibrator ($\vis'_c$), we need to transform $\ld$ into an
equivalent $\ud$ that has the same visibility amplitude at the observed
projected baseline and wavelength.  To do this, we calculated a correction
factor ($\rho_w=\ld/\ud$) based on $T_{\rm eff}$, metallicity
\citep[]{Nordstrom04}, and the transmission of the $K'$-filter
\citep[]{Davis00, Tango02}.  These correction factors ( Table~\ref{tbl-cal},
column~6) result in slightly smaller ($\approx 1.5 \%$) diameters for
equivalent uniform disks.

\subsection{Calibrated Visibility Amplitudes}

Because calibrator measurements were not made simultaneously with object
measurements, we linearly interpolated in time between the values of $\vis'_c$
immediately before and after each $\vis_o'$.  We then used the calibrator $\ud$
and the known values of $B$ and $\lambda_0$ to compute $\vis_c$. From $\vis'_c$
and $\vis_c$, we determined the instrumental efficiency ($\eta$) for visibility
measurement.  Typically values of $\eta$ ranged from $40$ to $50\%$ and varied
by only a few percent over the course of the night.  This can be attributed to
the stability of the system and habitual realignment of the optics before every
observation.  Finally, from eq.~\ref{eq:cal_vis}, we determined the calibrated
visibility amplitudes ($\vis_o$).  Table~\ref{tbl-visibilities} lists the
Modified Julian Dates (MJD), projected baselines, and visibility amplitudes
(with error estimates) associated with the mid-point time of observation for
each dwarf star.  Errors in calibrator angular size ($\S$\ref{sect-irfm}) and
visibility amplitudes were propagated to the calibrated visibility amplitude
error estimates.

\section{Angular Diameters and Radii}

We can determine the stellar radius from the angular diameter and parallax in
two ways.  First, we can assume that the stars are uniform disks and then fit
the fringe visibility as a function of baseline using eq.~\ref{eq:vis_ud}.
This estimate of $\ud$ is given in column~2 of Table~\ref{tbl-derived}.
However, we know that real stars are limb darkened (by a small amount in the
$K'$-band), so that their actual, limb darkened diameters will be slightly
larger than the uniform disk diameters (see Table~\ref{tbl-cal}).   In
principle, it is no harder to fit a limb darkened visibility curve to the
observations \citep[eq.~6]{Davis00}, but to do so we first need to establish
the stellar parameters in order to obtain the appropriate $K'$-band limb
darkening law from stellar atmosphere models.

We approached this problem as follows.  \citet[]{Claret00} has tabulated limb
darkening coefficients for the near-IR that are based upon solar metallicities.
We chose the coefficients calculated using the PHOENIX code for modeling
stellar atmospheres (Table 38 from \citealt[]{Claret00}).  The limb darkening
relations are listed as functions of $T_{\rm eff}$ and $\log g$.  We can
determine these parameters using an iterative scheme that is based upon
sequential improvements in the radius estimate.  We start with a stellar radius
estimate derived from the uniform disk angular diameter (column~2 of
Table~\ref{tbl-derived}) and the parallax from the NStars
database\footnote{http://nstars.nau.edu}.  NStars parallaxes are the weighted
means of all currently available parallaxes, including those of Hipparcos and
the Yale Parallax Catalog.  Then we use the Stefan-Boltzmann relation to find
$T_{\rm eff}$ from the bolometric luminosity and radius.  We determined the
bolometric luminosity from the absolute $K$ magnitude using a bolometric
correction based on the $I-K$ color index (derived from spectral energy
distribution fits of M-stars by \citealt{Leggett00}).   We used absolute $K$
magnitudes calculated from NStars parallaxes and 2MASS K magnitudes and $I-K$
color indices from \citet[]{Leggett92} to find the luminosities given in column
5 of Table~\ref{tbl-derived}, which were then used to find $T_{\rm eff}$ from
the assumed stellar radius.  We adopted $L_\sun = 3.86\times 10^{33}$
erg~s$^{-1}$ and $M_{bol\sun}=4.75$ in this calculation, and the scatter in the
bolometric correction relation results in a luminosity error of $\pm
0.022$~dex.  Next we used estimates of the stellar mass (column 4 of
Table~\ref{tbl-dwarfs}) with the assumed radius to find $\log g$.  Stellar
masses were estimated using the $K$-band mass--luminosity relation of
\citet[]{Delfosse00}, which have a typical error of $\pm 10\%$.  With $T_{\rm
eff}$ and $\log g$ set, we then found the limb darkening law from
\citet[]{Claret00} and made a least-squares error-weighted fit of the
limb-darkened visibility curve to the observations to obtain the limb darkened
angular size $\ld$.  

\begin{figure}
\begin{center}
\includegraphics[angle=90, scale=0.3]{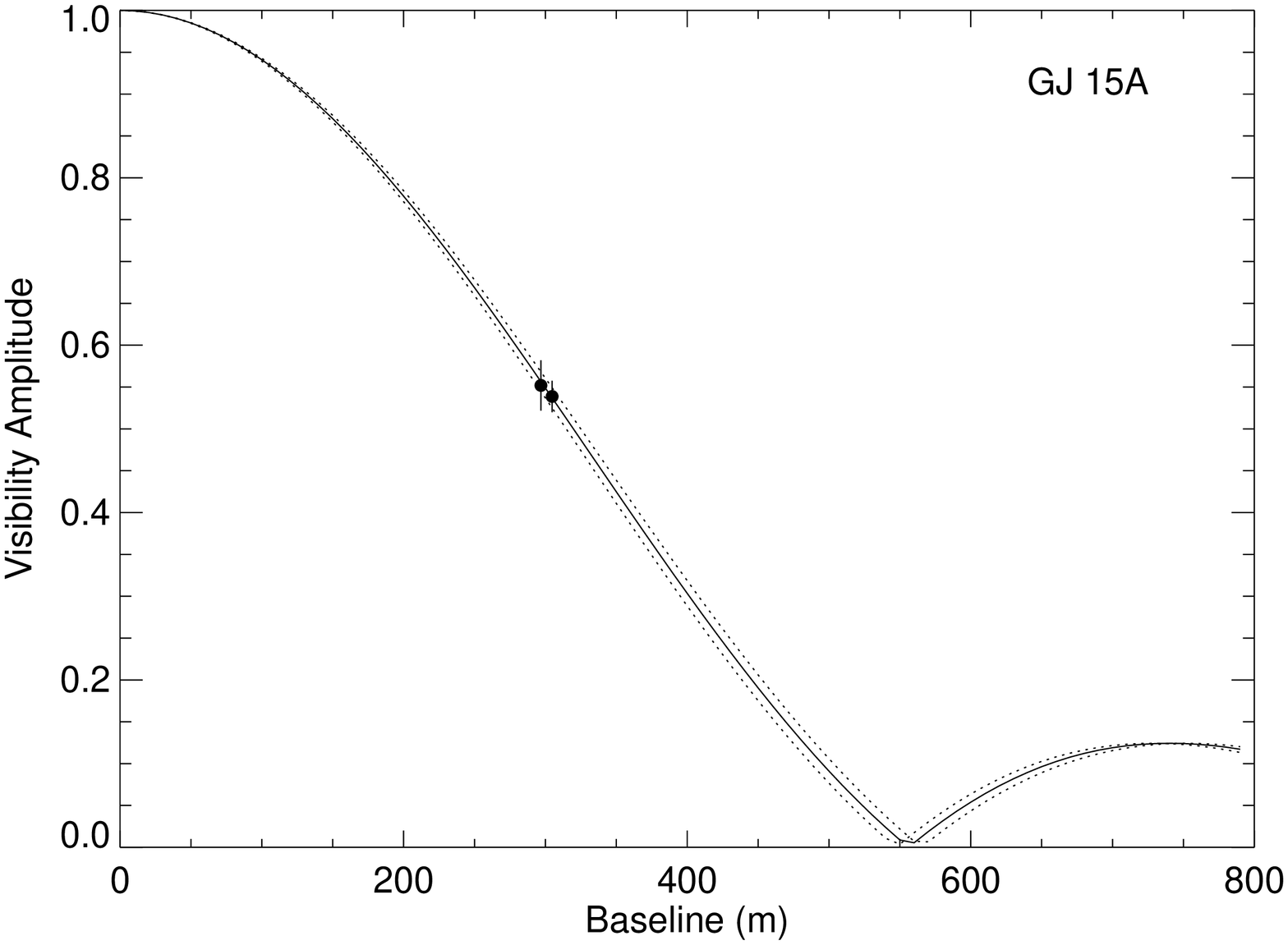}
\includegraphics[angle=90, scale=0.3]{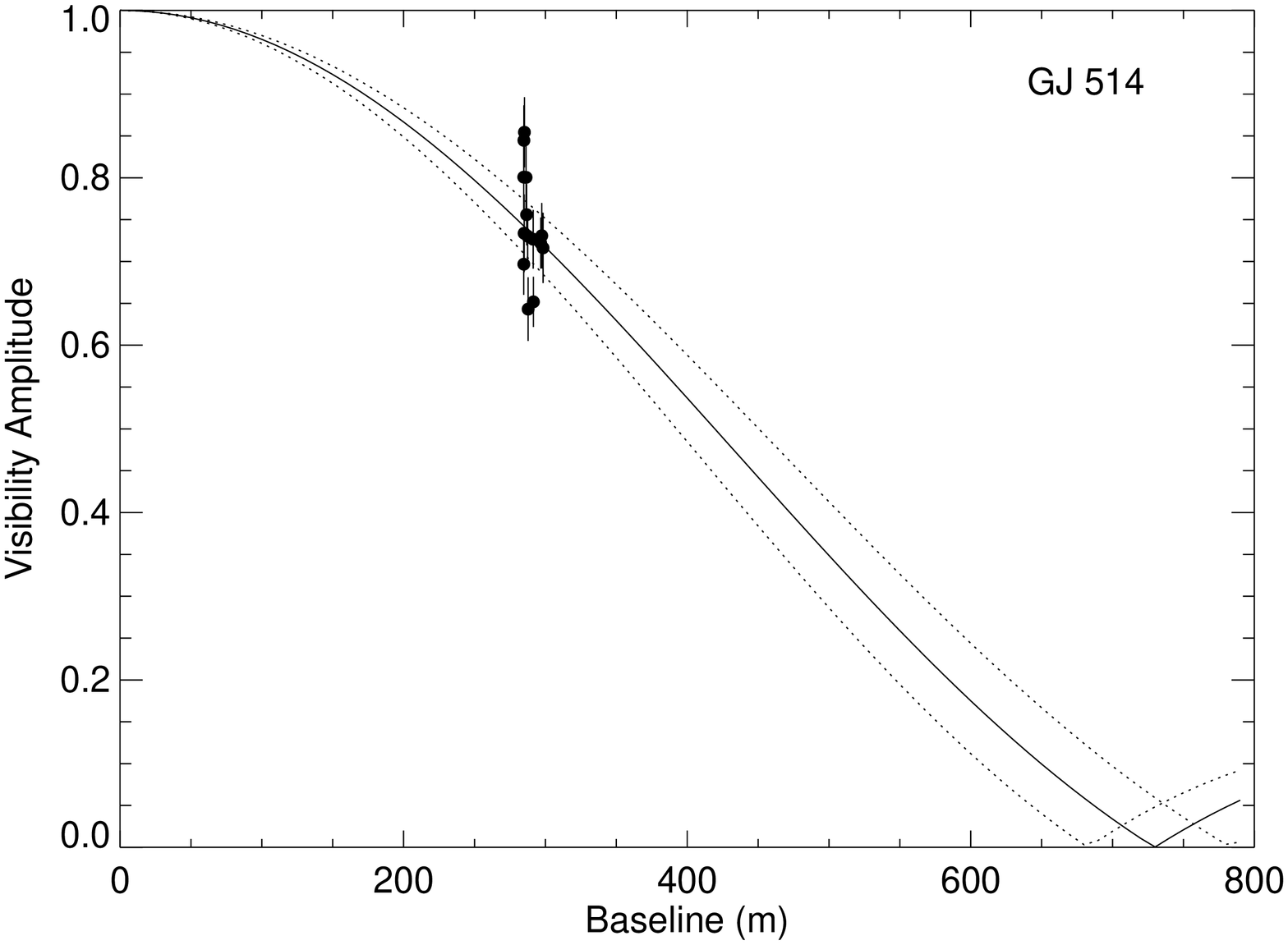}
\includegraphics[angle=90, scale=0.3]{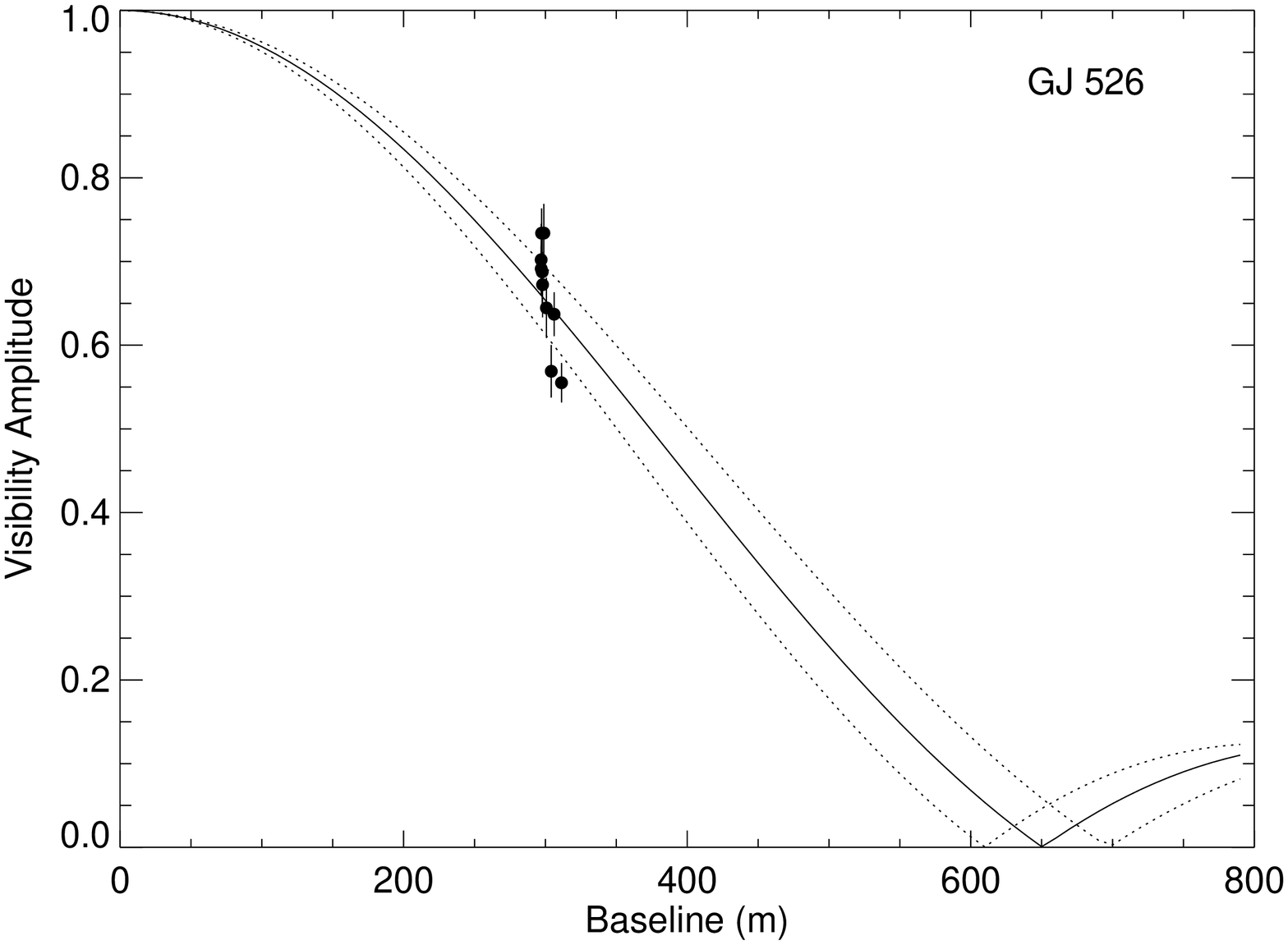}
\includegraphics[angle=90, scale=0.3]{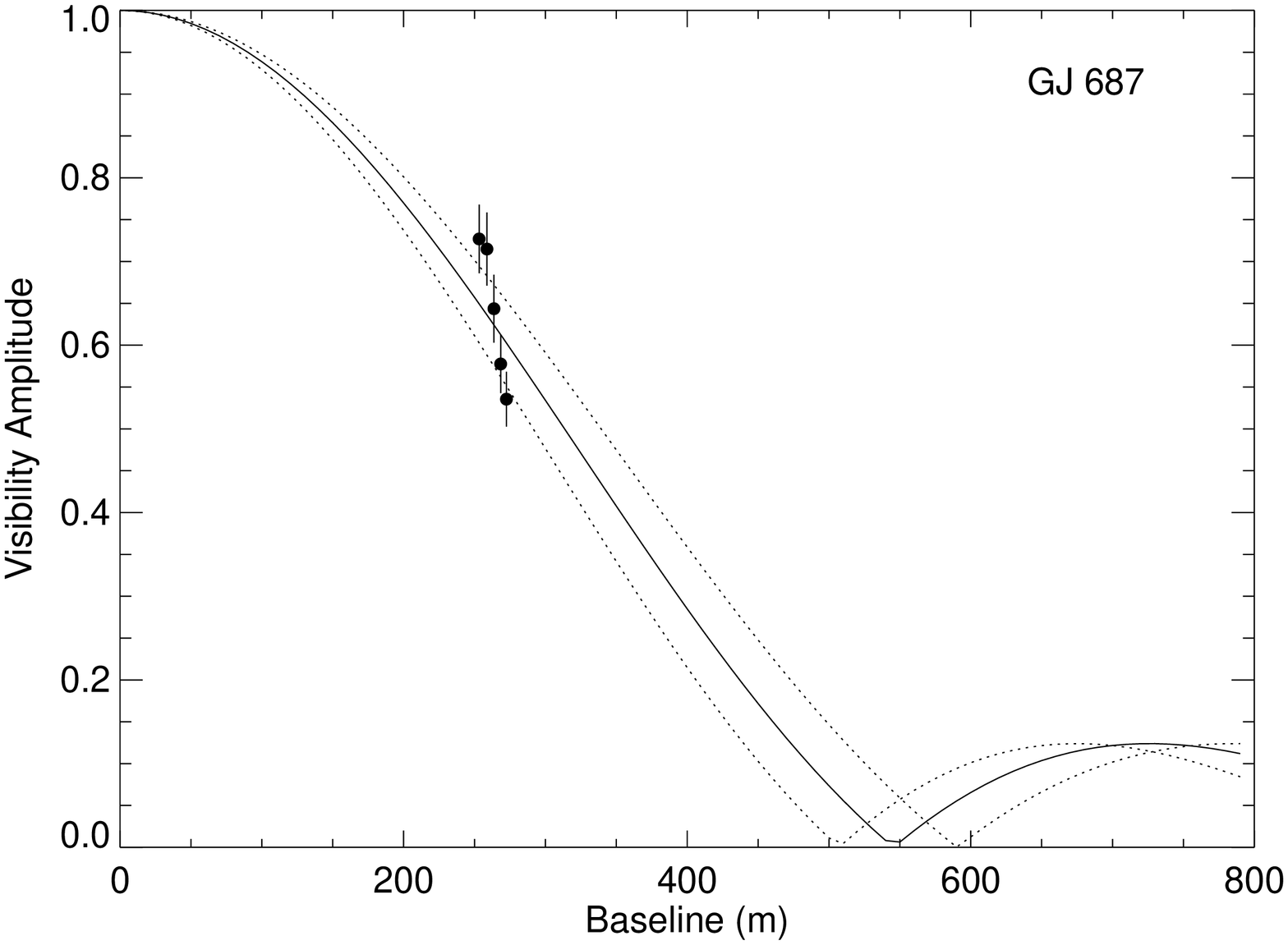}
\includegraphics[angle=90, scale=0.3]{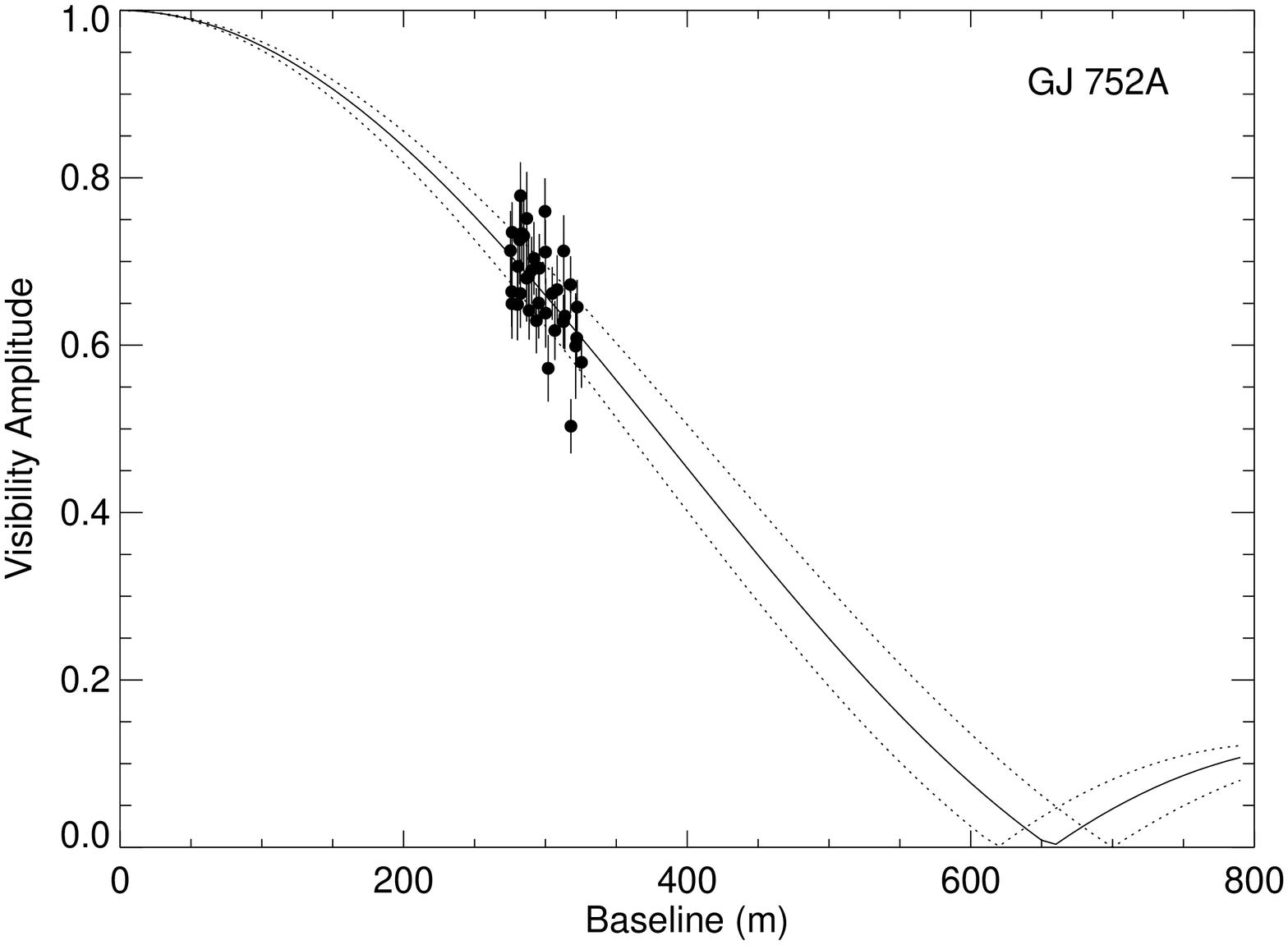}
\includegraphics[angle=90, scale=0.3]{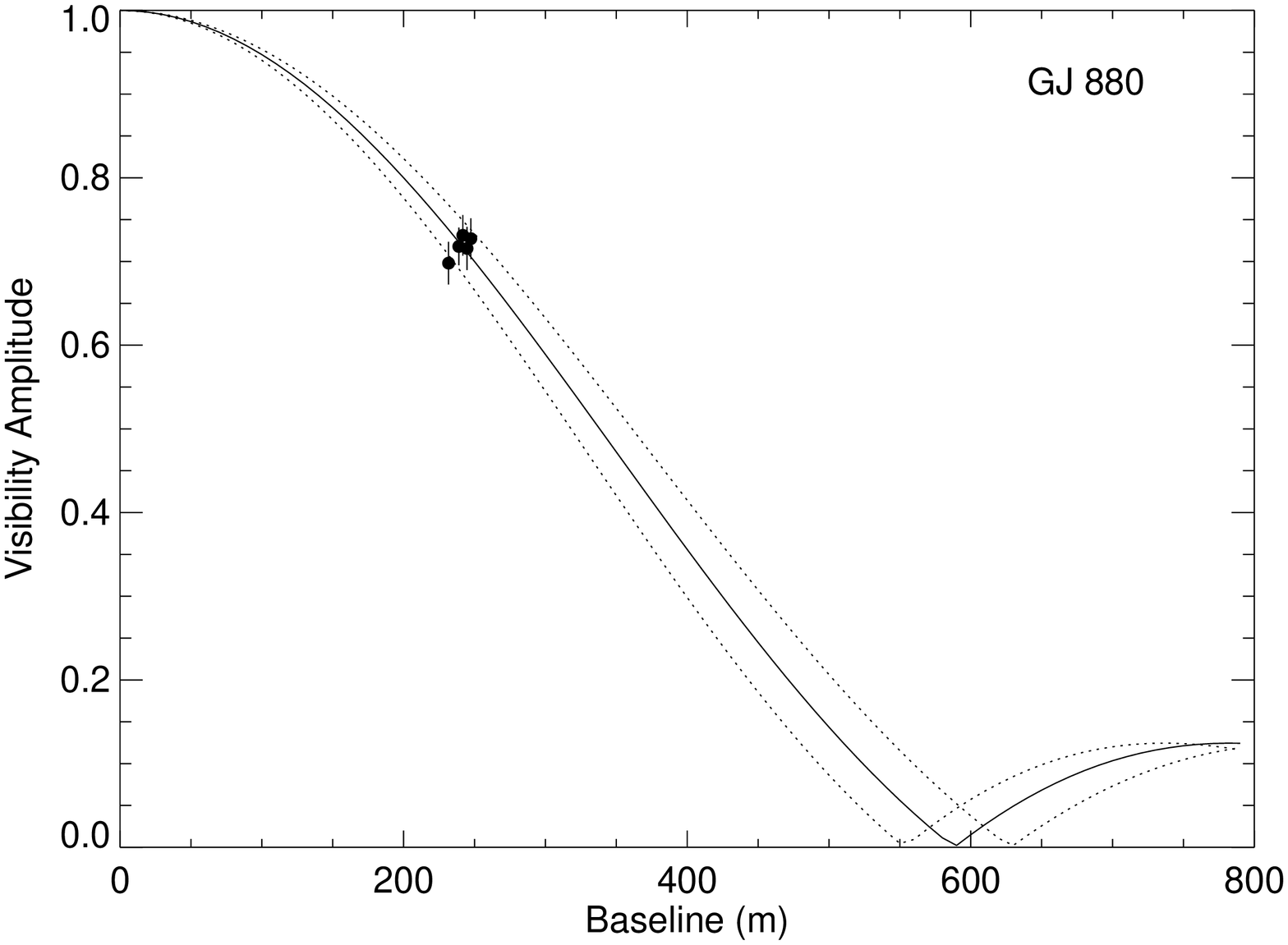}
\end{center}
\caption{Observed visibilities and fitted visibility curve (solid line) for a
limb darkened disk (using a $K$-band limb darkening law from
\citealt[]{Claret00} for the stellar parameters given in
Table~\ref{tbl-derived}).  Dotted lines represent the total error to the model
fit.\label{fig-ld_vis}}
\end{figure}

We repeated the process and revised the temperature and gravity estimates by
estimating a new radius from $\ld$ and parallax, and then fitted the
visibilities again with the revised limb darkened visibility curve.  In
practice this procedure converged (with negligible parameter differences
between iterations) in only two iterations because the IR limb darkening is
only slightly different from a uniform disk and is relatively insensitive to
the adopted temperature and gravity.  In fact, a 1-$\sigma$ change in the
adopted temperature and gravity has no detectable effect on $\ld$.   Our final
values of $\ld$, reduced chi-squared ($\chi^2_{\rm red}$), stellar radius,
$T_{\rm eff}$, and $\log g$ are listed in Table~\ref{tbl-derived} and the
fitted, limb darkened visibility curves are plotted with the observations in
Figure~\ref{fig-ld_vis}.  The fractional errors in the derived temperatures are
approximately one half the fractional errors in the radii or about $3\%$, and
the errors in $\log g$ amount to approximately $\pm 0.07$~dex (as derived from
the errors in fractional mass and radius).  Changing these parameters by one
standard deviation does not effect the radius determinations at the precision
we are reporting them.

The one inconsistency in this method is that we have relied on the solar
abundance atmospheric models from \citet[]{Claret00} to estimate the limb
darkening, while some of our targets are somewhat metal poor (column~3,
Table~\ref{tbl-dwarfs}).  It will be interesting to revisit our calculations
when limb darkening results for metal poor atmospheres are eventually
developed, but we doubt that our radius results will change significantly
because the limb darkening corrections are small.

\begin{figure}
\begin{center}
\includegraphics[angle=90, scale=0.6]{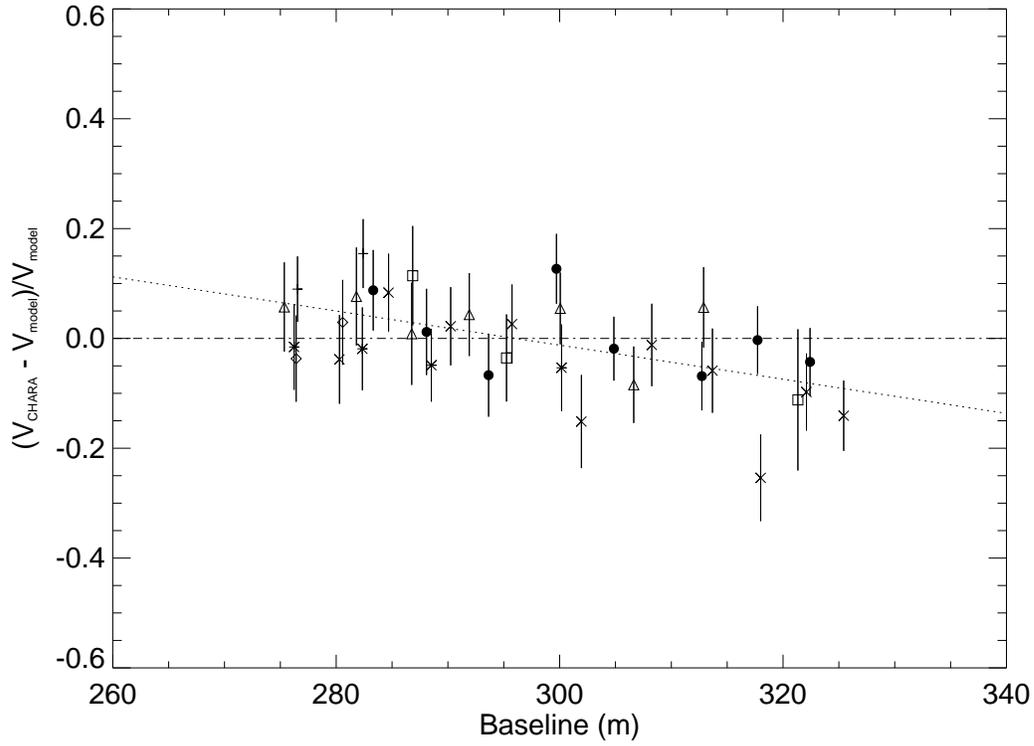}
\caption{Fractional deviation between the observed visibilities and fitted
visibility curve for a limb darkened disk (same as in Fig.~\ref{fig-ld_vis})
for GJ 752A.  The dotted line represents a best fit regression line.  Symbols indicated data taken on different nights (
$+$		= 2004 Jun 5,
$\ast$		= 2004 Jun 6,
$\Diamond$	= 2004 Jun 8,
$\bigtriangleup$= 2004 Jun 11,
$\Box$		= 2004 Jun 12,
$\times$	= 2004 Jun 13,
$\bullet$	= 2004 Jun 14
).\label{fig-gj752a}}
\end{center}
\end{figure}

We found that the values of $\chi^2_{\rm red}$ (given in column 4 of Table
\ref{tbl-derived}) exceeded the expected value of unity for all the targets
except GJ~15A and GJ~880 (the small value for GJ~15A probably results from our
sample of two measurements).  Thus, the internal visibility amplitude errors
associated with an individual data set underestimate the full error budget, and
consequently, we used the observed scatter from the fits to rescale the
minimum, best-fit value of $\chi^2_{\rm red}$ to unity.  We added a 6\% noise
floor to account for the night-to-night fluctuation in angular diameter
measurement, as calculated from objects with multi-night observations.  This
term was added in quadrature to the statistical error.  We show in
Figure~\ref{fig-gj752a} the distribution of the fractional deviations from the
fit for GJ~752A, the target with the most observations.  We also plot a linear
regression fit of the residuals that has a non-zero slope which is set by a few
outlying points from two nights (2004 June 13-14).  The 6\% noise floor
also accounts for this slight linear trend seen in these commissioning
observations.

We also determined effective temperatures by a direct comparison of the
observed and predicted fluxes in the $K$-band based upon the observed angular
diameters.  We used the 2MASS $K_s$ band magnitudes and the adopted flux
zero-point from \citet[]{Cohen03} to form the following relation:
\begin{equation} 
K_s = -2.5 \log F_\lambda - 5 \log \theta_{LD} + 17.157
\end{equation} 
where $F_\lambda$ is the model flux (erg~cm$^{-2}$~s$^{-1}$~\AA $^{-1}$)
averaged over the 2MASS $K_s$ filter response \citep[]{Cohen03} and $\ld$ is
the limb-darkened angular diameter (mas).  The model fluxes were taken from the
PHOENIX atmosphere code of \citet{Brott05}\footnote{
ftp://ftp.hs.uni-hamburg.de/pub/outgoing/phoenix/GAIA/} and these are
primarily functions of effective temperature (although we did interpolate in
these models for the appropriate gravity and metallicity of each target).  We
used our values of $\ld$ from Table~\ref{tbl-derived} to find the estimates of
$F_\lambda (T_{\rm eff})$ and hence effective temperature that are listed in
column~8 of Table~\ref{tbl-derived} (under the heading $T_{\rm eff}$(2MASS)).
The temperatures agree well with those from the bolometric correction method,
$T_{\rm eff}$(BC), discussed above.   Note that the adopted limb-darkened
diameters themselves depend on the assumed temperature through the limb
darkening coefficients, but since the resulting temperatures are so similar and
the limb darkening is a minor effect, this approximation has a negligible
impact on the temperature derived this way.

Finally we note that we find no evidence in our data that any target has a
close binary companion that could affect the radius estimates.  The targets are
all radial velocity constant (r.m.s.\ $<100$ m~s$^{-1}$) according to the
spectroscopic survey of \citet[]{Nidever02}.  Optical speckle interferometry by
\citet{McAlister87} and \citet{Balega99} indicates that GJ~687 has a companion
at a separation of $0\farcs3$, which is too wide to influence our observations
(except as a source of incoherent light that might reduce the measured
visibility).  However, near-infrared speckle observations \citep[]{Leinert97}
and high resolution HST/NICMOS imaging (T.\ J.\ Henry, private communication)
show no evidence of a companion at such a separation.  We examined the mean
fringe envelopes for each set of scans of this star and found no detectable
companion with $\Delta K_s < 2$ in the separation range 12 -- 70 mas.  Clearly
more observations are required to settle the question about a companion to
GJ~687, but we tentatively assume that the measured visibility is dominated by
the photospheric disk of the bright M-dwarf.

\section{Mass--Radius Relation}

\begin{figure}
\begin{center}
\includegraphics[angle=90, scale=0.6]{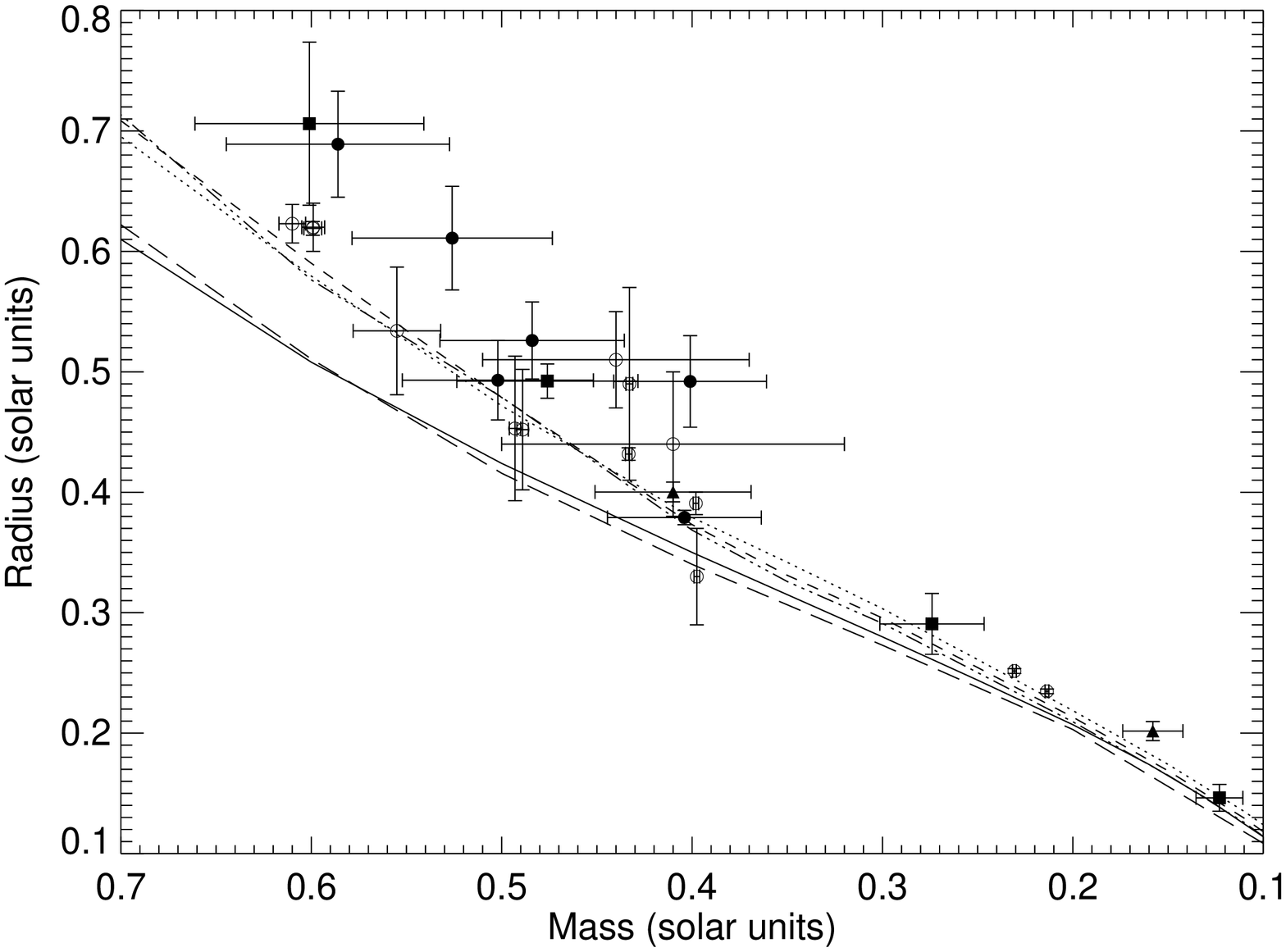}
\caption{The mass--radius relation for low-mass dwarfs measured by long-baseline
interferometry (filled symbols) and spectrophotometry of eclipsing binaries
(open circles, see references in \S~\ref{sect-intro}).  The interferometry data
included are from this paper (circles), PTI \citep[triangles]{Lane01},
and VLTI \citep[squares]{Segransan03}.  The lines represent models from
\citet[]{Chabrier97} for different metallicities ($\cdot\cdot\cdot\cdot~\!\!$
for [M/H]$=0.0$; - - - - for [M/H]$=-0.5$; $-\cdot\cdot\cdot~\!\!$ for
[M/H]$=-1.0$) and \citet[]{Siess97} for similar metallicities (--- for
[M/H]$=0.0$; -- -- -- for [M/H]$=-0.3$).
\label{fig-mass_radius}}
\end{center} \end{figure}

In an effort to further constrain the poorly populated mass--radius relation for
low-mass stars, we started by comparing our results with other low-mass stellar
radii directly measured with long-baseline interferometers \citep[]{Lane01,
Segransan03}.  All masses were derived from the same $K$-band mass--luminosity
relation of \citet[]{Delfosse00}, and we adopted a 10\% error to account for
photometric and empirical fitting errors.  Note that the mass derived from the
absolute $K$ magnitude should have little or no dependence upon the stellar
metallicity (see Fig.\ 3b of \citealt{Baraffe98}).  \citet[]{Lane01} also
measured the angular diameter of GJ 15A and their value of $\ud=0.985\pm0.05$
mas agrees with our value $\ud=0.976\pm0.016$ mas.  However, we adopted a less
severe limb-darkening correction resulting in a smaller linear diameter.  In
our mass--radius plot (Figure~\ref{fig-mass_radius}), we only show our data
point for GJ 15A.  Eclipsing binary star measurements are also shown in
Figure~\ref{fig-mass_radius} as open circles, and it should be noted that some
of the errors bars are smaller than the symbol size.  The theoretical models of
\citet[]{Chabrier97} for [M/H] $= 0$, $-0.5$, and $-1.0$
are shown as the lines.  Additionally, the models of \citet[]{Siess97} for
[M/H] $= 0$, $-0.3$ (assuming Z$_\odot=0.02$) are plotted to show the range of
model uncertainty.  According to these models, metallicity should have little
effect on radius for a given mass star.  However, there is either a systematic
effect in the data or the models are underestimating the sizes of the stars for
a given mass.  We doubt the former possibility because this effect is seen in
data from multiple instruments and is present in both the interferometric and
binary results.  Indeed, the larger than expected radii have already been noted
in several investigations \citep[]{Leggett00, Segransan03, LP05}.

\citet[]{Mullan01} argue that larger radii could result from pervasive magnetic
fields that could alter the interior structure and push the occurrence of
completely convective interiors to stars of lower effective temperatures.  They
suggest that M-stars with active magnetic fields will have larger radii than
predicted by standard models.  However, there is no evidence that the M-dwarfs
in our sample have any special magnetic properties.  All are slow rotators
\citep[]{Delfosse98}, have moderate X-ray coronal emission \citep[]{Schmitt95,
Huensch99}, have normal chromospheric H$\alpha$ lines \citep[]{Gizis02}, and
show no evidence of photometric variability in {\it Hipparcos} photometry
\citep[]{Perryman97}.   Thus, these stars do not have the exceptionally strong
magnetic fields that would lead to larger radii according to the models
of \citeauthor{Mullan01}.

\begin{figure}
\begin{center}
\includegraphics[angle=90, scale=0.6]{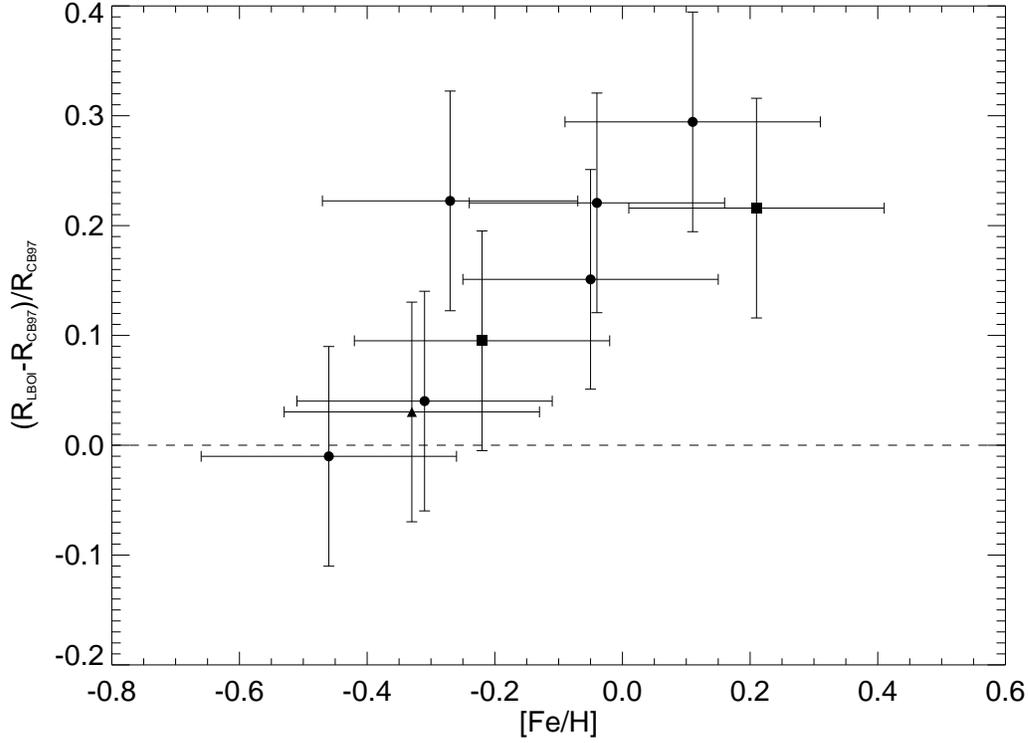}
\caption{Fractional deviation between the radii measured through long baseline
optical interferometry (designated LBOI) and from the model predictions for
stellar radius from \citet[designated CB97]{Chabrier97} plotted as a function
of metallicity.  The symbols represent the same observational groups given in
Fig.~\ref{fig-mass_radius}.  The representative errors are $\pm 0.2$ dex in
[Fe/H] and $\pm 0.1$ in fractional deviation of the radius (due to $10\%$
errors in the mass estimates).\label{fig-met_radius}}
\end{center}
\end{figure}

On the other hand, our sample stars do span a significant range in metallicity,
and metallicity will play a role in the internal structure of M-stars.  To
investigate this possibility, we plot  in Figure~\ref{fig-met_radius} the
fractional deviation from the [M/H]=0 model prediction as a function of
metallicity.  The metallicities for most stars are taken from
\citet[]{Bonfils05} except for those for GJ~191, GJ~205, GJ~887
\citep[]{Woolf05}, and GJ 699 \citep[]{Segransan03}.  Efforts to derive
accurate metallicities for M dwarfs are in the nascent phase at this time, so
the metallicity values should be considered preliminary, although the general
trends are likely reliable.  We see that the observed radii are approximately
consistent with predictions among the metal-poor stars, but the radii become
larger than predicted as the metallicity increases.  The same conclusion can be
drawn from the results of \citet[]{Leggett00} who show that the radii derived
from spectral fits are systematically larger among stars with higher
metallicity when plotted in a $(T_{\rm eff}, R)$ diagram (see their Fig.~13).
Because metallicity is closely related to stellar opacity, we suspect that the
current generation of models for the interiors and atmospheres of M-stars is
missing some opacity component that results in larger radii for stars of higher
metallicity.  We are planning to expand this investigation to other targets
with a wider range in metallicity in order to explore this connection further.
This work will include more spatial coverage via different baselines and
observations at shorter wavelengths to obtain measurements farther along the
visibility curves.

\acknowledgments

We would like to thank PJ Goldfinger, Mandy Henderson, Steve Golden, and Bob
Cadman for their tireless efforts at the Array.
Construction funding for the CHARA Array has been provided by the National
Science Foundation through grant AST 94-14449, the W. M. Keck Foundation, the
David and Lucile Packard Foundation, and by Georgia State University.  Science
operations at the Array are supported by the National Science Foundation
through NSF Grant AST-0307562 and by Georgia State University through the
College of Arts and Sciences and the Office of the Vice President for Research.
Financial support for DHB was provided by the National Science Foundation
through grant AST--0205297.
This research has made use of the SIMBAD and VizieR \citep[]{Ochsenbein00}
databases, operated at CDS, Strasbourg, France.  
This publication made use of data products from the Two Micron All Sky Survey,
which is a joint project of the University of Massachusetts and the Infrared
Processing and Analysis Center/California Institute of Technology, funded by
the National Aeronautics and Space Administration and the National Science
Foundation.


\begin{deluxetable}{cccclc}
\tabletypesize{\scriptsize}
\tablecaption{Observations\label{tbl-obs}}
\tablewidth{0pt}
\tablehead{
	\colhead{GJ} &  
	\colhead{LHS} &  
	\colhead{Calibrator} &  
	\colhead{Baseline} & 
	\colhead{Date} & 
	\colhead{No. of}\\
	\colhead{No.} &
	\colhead{No.} &
	\colhead{Name} &  
	\colhead{(m)} &
	\colhead{(UT)} &
	\colhead{Observations}
}
\startdata
15A	& 3	& HD 2952	& W1/E1 (314 m) & 2004 Oct \phn 9 & \phn 2  \\
\tableline
514	& 352	& HD 119550	& S1/E1 (331 m) & 2004 Jun     11 & \phn 4  \\
	& 	&		&		& 2004 Jun     12 & \phn 4  \\
	& 	&		&		& 2004 Jun     14 & \phn 6  \\
\tableline
526	& 47	& HD 119550	& S1/E1 (331 m) & 2004 Jun \phn 6 & \phn 3  \\
	& 	&		&		& 2004 Jun \phn 7 & \phn 1  \\
	& 	&		&		& 2004 Jun \phn 8 & \phn 1  \\
	& 	&		&		& 2004 Jun     13 & \phn 5  \\
\tableline
687	& 450	& HD 151541	& S1/E1 (331 m) & 2004 Jun     26 & \phn 5  \\
\tableline
752A	& 473	& HD 182101	& S1/E1 (331 m) & 2004 Jun \phn 5 & \phn 2  \\
	& 	&		&		& 2004 Jun \phn 6 & \phn 4  \\
	& 	&		&		& 2004 Jun \phn 8 & \phn 2  \\
	& 	&		&		& 2004 Jun     11 & \phn 7  \\
	& 	&		&		& 2004 Jun     12 & \phn 3  \\
	& 	&		&		& 2004 Jun     13 &      10 \\
	& 	&		&		& 2004 Jun     14 & \phn 8  \\
\tableline
880	& 533	& HD 218261	& S2/W1 (249 m)	& 2003 Dec     16 & \phn 5  \\
\enddata
\end{deluxetable}

\begin{deluxetable}{ccccccc}
\tabletypesize{\scriptsize}
\tablecaption{Calibrators\label{tbl-cal}}
\tablewidth{0pt}
\tablehead{
	\colhead{HD} & 
	\colhead{Spectral} &
	\colhead{$\log(T_{\rm eff})$\tablenotemark{a}} &
	\colhead{} &
	\colhead{$\theta_{LD}$} & 
	\colhead{} & 
	\colhead{$\theta_{UD}$}\\
	\colhead{No.} &
	\colhead{Classification} &
	\colhead{(K)} &
	\colhead{[Fe/H]\tablenotemark{a}} &
	\colhead{(mas)} &
	\colhead{$\rho_w$} &
	\colhead{(mas)}\\
}
\startdata
119550	& G2 V\tablenotemark{b}   & $3.755$ & $-0.07$    & $0.363 \pm 0.015$ & $1.0148$ & $0.358 \pm 0.015$ \\
151541	& K1 V\tablenotemark{c}   & $3.719$ & $-0.36$    & $0.326 \pm 0.013$ & $1.0165$ & $0.321 \pm 0.013$ \\
182101	& F6 V\tablenotemark{d}   & $3.799$ & $-0.42$    & $0.367 \pm 0.017$ & $1.0120$ & $0.363 \pm 0.017$ \\
218261	& F8.5 V\tablenotemark{e} & $3.799$ & \phs$0.00$ & $0.387 \pm 0.021$ & $1.0139$ & $0.382 \pm 0.021$ \\
\enddata
\tablenotetext{a}{\citet[]{Nordstrom04}}
\tablenotetext{b}{\citet[]{Harlan69}}
\tablenotetext{c}{\citet[]{Cowley67}}
\tablenotetext{d}{\citet[]{Bidelman57}}
\tablenotetext{e}{\citet[]{Gray01}}
\end{deluxetable}

\begin{deluxetable}{cccc}
\tabletypesize{\scriptsize}
\tablecaption{Calibrated $K'$ Visibilities\label{tbl-visibilities}}
\tablewidth{0pt}
\tablehead{
	\colhead{Object} &
	\colhead{Date} &
	\colhead{Baseline} &
	\colhead{}\\
	\colhead{Name} &
	\colhead{(MJD)} &
	\colhead{(m)} &
	\colhead{Visibility}
}
\startdata
GJ 15A  & 53287.364 & 304.7 & $0.290 \pm 0.013$\tablenotemark{a} \\
	& 53287.390 & 296.8 & $0.305 \pm 0.021$\tablenotemark{a} \\
\tableline
GJ 514  & 53167.188 & 298.3 & $0.716 \pm 0.042$ \\
        & 53167.230 & 287.8 & $0.643 \pm 0.038$ \\
        & 53167.259 & 284.8 & $0.697 \pm 0.037$ \\
        & 53167.279 & 285.0 & $0.734 \pm 0.046$ \\
        & 53168.188 & 297.4 & $0.731 \pm 0.039$ \\
        & 53168.210 & 291.4 & $0.727 \pm 0.035$ \\
        & 53168.234 & 286.7 & $0.756 \pm 0.037$ \\
        & 53168.255 & 284.8 & $0.845 \pm 0.042$ \\
        & 53170.185 & 296.7 & $0.722 \pm 0.030$ \\
        & 53170.204 & 291.5 & $0.652 \pm 0.030$ \\
        & 53170.224 & 287.5 & $0.730 \pm 0.033$ \\
        & 53170.243 & 285.2 & $0.854 \pm 0.042$ \\
        & 53170.265 & 284.7 & $0.801 \pm 0.038$ \\
        & 53170.285 & 286.4 & $0.801 \pm 0.045$ \\
\tableline
GJ 526  & 53162.272 & 297.7 & $0.688 \pm 0.043$ \\
        & 53162.291 & 297.0 & $0.702 \pm 0.043$ \\
        & 53162.313 & 298.0 & $0.672 \pm 0.039$ \\
        & 53163.245 & 300.7 & $0.645 \pm 0.036$ \\
        & 53164.224 & 304.1 & $0.569 \pm 0.031$ \\
        & 53169.180 & 311.4 & $0.555 \pm 0.024$ \\
        & 53169.201 & 306.1 & $0.637 \pm 0.026$ \\
        & 53169.241 & 298.9 & $0.734 \pm 0.035$ \\
        & 53169.259 & 297.3 & $0.734 \pm 0.030$ \\
        & 53169.278 & 297.0 & $0.691 \pm 0.028$ \\
\tableline
GJ 687  & 53182.186 & 253.3 & $0.727 \pm 0.041$ \\
	& 53182.200 & 258.8 & $0.715 \pm 0.044$ \\
	& 53182.213 & 263.7 & $0.644 \pm 0.041$ \\
	& 53182.227 & 268.4 & $0.578 \pm 0.035$ \\
	& 53182.241 & 272.5 & $0.535 \pm 0.033$ \\
\tableline
GJ 752A & 53161.452 & 282.4 & $0.779 \pm 0.040$ \\
        & 53161.470 & 276.5 & $0.735 \pm 0.036$ \\
        & 53162.405 & 300.2 & $0.638 \pm 0.041$ \\
        & 53162.433 & 288.5 & $0.641 \pm 0.035$ \\
        & 53162.449 & 282.3 & $0.662 \pm 0.041$ \\
        & 53162.468 & 276.2 & $0.664 \pm 0.042$ \\
        & 53164.449 & 280.6 & $0.694 \pm 0.044$ \\
        & 53164.462 & 276.4 & $0.650 \pm 0.042$ \\
        & 53167.359 & 312.9 & $0.713 \pm 0.043$ \\
        & 53167.376 & 306.6 & $0.618 \pm 0.035$ \\
        & 53167.391 & 300.1 & $0.711 \pm 0.038$ \\
        & 53167.411 & 291.9 & $0.704 \pm 0.043$ \\
        & 53167.424 & 286.8 & $0.680 \pm 0.052$ \\
        & 53167.437 & 281.8 & $0.726 \pm 0.053$ \\
        & 53167.457 & 275.4 & $0.713 \pm 0.047$ \\
        & 53168.332 & 321.3 & $0.599 \pm 0.063$ \\
        & 53168.400 & 295.2 & $0.650 \pm 0.042$ \\
        & 53168.421 & 286.8 & $0.751 \pm 0.056$ \\
        & 53169.313 & 325.4 & $0.580 \pm 0.030$ \\
        & 53169.326 & 322.1 & $0.609 \pm 0.035$ \\
        & 53169.339 & 318.0 & $0.503 \pm 0.032$ \\
        & 53169.352 & 313.7 & $0.635 \pm 0.040$ \\
        & 53169.366 & 308.3 & $0.666 \pm 0.041$ \\
        & 53169.381 & 301.9 & $0.572 \pm 0.040$ \\
        & 53169.396 & 295.7 & $0.692 \pm 0.041$ \\
        & 53169.410 & 290.2 & $0.689 \pm 0.040$ \\
        & 53169.424 & 284.7 & $0.731 \pm 0.042$ \\
        & 53169.436 & 280.3 & $0.649 \pm 0.043$ \\
        & 53170.322 & 322.4 & $0.646 \pm 0.033$ \\
        & 53170.338 & 317.7 & $0.672 \pm 0.034$ \\
        & 53170.352 & 312.7 & $0.628 \pm 0.032$ \\
        & 53170.372 & 304.9 & $0.662 \pm 0.032$ \\
        & 53170.384 & 299.7 & $0.760 \pm 0.040$ \\
        & 53170.399 & 293.6 & $0.629 \pm 0.039$ \\
        & 53170.412 & 288.1 & $0.682 \pm 0.044$ \\
        & 53170.425 & 283.3 & $0.734 \pm 0.044$ \\

\tableline
GJ 880  & 52928.301 & 247.4 & $0.727 \pm 0.024$ \\
        & 52928.315 & 244.7 & $0.716 \pm 0.026$ \\
        & 52928.326 & 241.7 & $0.731 \pm 0.024$ \\
        & 52928.334 & 238.9 & $0.718 \pm 0.022$ \\
        & 52928.351 & 231.7 & $0.698 \pm 0.025$ \\
\enddata
\tablenotetext{a}{Visibilities are given as $\vis^2$}
\end{deluxetable}

\begin{deluxetable}{lcccccccc}
\tabletypesize{\scriptsize}
\tablecaption{Derived Stellar Parameters\label{tbl-derived}}
\tablewidth{0pt}
\tablehead{
	\colhead{Object} & 
	\colhead{$\theta_{UD}$} &
	\colhead{$\theta_{LD}$} &
	\colhead{$\chi^2_{\rm red}$} &
	\colhead{$\log L_{bol} $} &
	\colhead{Radius} &
	\colhead{$T_{\rm eff}$(BC)} &
	\colhead{$T_{\rm eff}$(2MASS)} &
	\colhead{$\log g$} \\
	\colhead{Name}&
	\colhead{(mas)}&
	\colhead{(mas)}&
	\colhead{} &
	\colhead{(erg s$^{-1}$)}&
	\colhead{($R_\Sun$)}&
	\colhead{(K)}&
	\colhead{(K)}&
	\colhead{(cm s$^{-2}$)}
} 
\startdata
GJ 15A	& $0.976 \pm 0.016$ & $0.988 \pm 0.016$ & $0.06$ & $31.99$ & $0.379 \pm 0.006$ & 3747 & $3730 \pm  \phn 49$ & 4.89 \\
GJ 514	& $0.740 \pm 0.044$ & $0.753 \pm 0.052$ & $2.50$ & $32.22$ & $0.611 \pm 0.043$ & 3377 & $3243 \pm 160$ & 4.59 \\
GJ 526	& $0.830 \pm 0.050$ & $0.845 \pm 0.057$ & $3.41$ & $32.18$ & $0.493 \pm 0.033$ & 3662 & $3636 \pm 163$ & 4.75 \\
GJ 687	& $0.990 \pm 0.059$ & $1.009 \pm 0.077$ & $3.06$ & $31.91$ & $0.492 \pm 0.038$ & 3142 & $3095 \pm 107$ & 4.66 \\
GJ 752A	& $0.822 \pm 0.049$ & $0.836 \pm 0.051$ & $1.49$ & $32.10$ & $0.526 \pm 0.032$ & 3390 & $3368 \pm 137$ & 4.68 \\
GJ 880	& $0.918 \pm 0.055$ & $0.934 \pm 0.059$ & $0.92$ & $32.32$ & $0.689 \pm 0.044$ & 3373 & $3277 \pm \phn 93$ & 4.53 \\
\enddata
\end{deluxetable}

\begin{deluxetable}{lclll}
\tabletypesize{\scriptsize}
\tablecaption{Adopted Stellar Parameters\label{tbl-dwarfs}}
\tablewidth{0pt}
\tablehead{
	\colhead{Object} & 
	\colhead{Spectral} &
	\colhead{} &
        \colhead{Mass\tablenotemark{c}} &
	\colhead{Parallax\tablenotemark{d}} \\
	\colhead{Name}&
	\colhead{Classification\tablenotemark{a}} &
	\colhead{[Fe/H]\tablenotemark{b}} &
        \colhead{($\mathcal{M}_\Sun$)} &
	\colhead{(arcsec)} \\
	
} 
\startdata
GJ 15A	& M1.5 V & $-0.46$ & 0.404 & 0.28059 $\pm$ 0.00095 \\ 
GJ 514	& M1.0 V & $-0.27$ & 0.526 & 0.13257 $\pm$ 0.00118 \\ 
GJ 526	& M1.5 V & $-0.31$ & 0.502 & 0.18421 $\pm$ 0.00116 \\ 
GJ 687	& M3.0 V & $+0.11$ & 0.401 & 0.22049 $\pm$ 0.00082 \\ 
GJ 752A	& M3.0 V & $-0.05$ & 0.484 & 0.17101 $\pm$ 0.00062 \\ 
GJ 880	& M1.5 V & $-0.04$ & 0.586 & 0.14579 $\pm$ 0.00113 \\ 
\enddata
\tablenotetext{a}{From \citet[]{Henry94}.}
\tablenotetext{b}{From \citet[]{Bonfils05}.  The typical error in [Fe/H] is $\pm 0.2$ dex.}
\tablenotetext{c}{From \citet[]{Delfosse00} and we adopted a 10\% error in
mass}
\tablenotetext{d}{NStars database, http://nstars.nau.edu}
\end{deluxetable}
\clearpage

\end{document}